\documentclass[aps,twocolumn,floatfix]{revtex4}
\usepackage{epsfig}
\usepackage{graphicx}
\usepackage{dcolumn}
\usepackage{natbib}

\def\dj{\hbox{d\kern-0,347em \vrule width0,3em height1,252ex
depth-1,21ex \kern0,051em}}
 
\begin{document}
  
\title{Crack roughness and avalanche precursors in the random fuse model}

\author{Stefano Zapperi}
\affiliation{INFM UdR Roma 1 and SMC, Dipartimento di Fisica,
Universit\`a "La Sapienza", P.le A. Moro 2, 00185 Roma, Italy}
\author{Phani Kumar V.V. Nukala}
\affiliation{Computer Science and Mathematics Division, 
Oak Ridge National Laboratory, Oak Ridge, TN 37831-6359, USA}
\author{Sr{\dj}an \v{S}imunovi\'{c}}
\affiliation{Computer Science and Mathematics Division, 
Oak Ridge National Laboratory, Oak Ridge, TN 37831-6359, USA}
 
\begin{abstract}
We analyze the scaling of the crack roughness and of avalanche precursors
in the two dimensional random fuse model by numerical simulations, employing large
system sizes and extensive sample averaging. We find that the crack roughness
exhibits anomalous scaling, as recently observed in experiments.  The roughness
exponents ($\zeta$, $\zeta_{loc}$) and the global width distributions
are found to be universal with respect to the lattice geometry.
Failure is preceded by avalanche precursors whose distribution follows
a power law up to a cutoff size. While the characteristic avalanche size scales 
as $s_0 \sim L^D$, with a universal fractal dimension $D$, 
the distribution exponent $\tau$ differs slightly
for triangular and diamond lattices and, in both cases, it is larger
than the mean-field (fiber bundle) value $\tau=5/2$.
\end{abstract}

%\PACS{46.50.+a, 64.60.Ak}
\maketitle

\section{introduction}

Understanding the scaling properties of fracture in disordered media
represents an intriguing theoretical problem with some technological 
implications \cite{breakdown}.  Experiments have shown that in several materials
under different loading conditions the fracture surface is
self-affine \cite{man} and the out of plane roughness exponent 
displays a universal value irrespective of the
material studied \cite{bouch}. In particular, experiments have been
done in metals \cite{metals}, glass \cite{glass}, rocks \cite{rocks}
and ceramics \cite{cera}, covering both ductile and brittle
materials. 

It was later shown that the roughness exponent conventionally measured
describes only the local properties, while the fracture surface 
instead exhibits anomalous scaling \cite{anomalous}: the {\it global} exponent describing the
scaling of the crack width with the sample size is larger than
the local exponent measured on a single sample \cite{exp-ano}. It is thus necessary to
define two roughness exponents a global ($\zeta$) and a local $\zeta_{loc}$.
Only the latter appears to be universal with a value $\zeta_{loc} \simeq 0.8$ \cite{bouch}.
For the purpose of this paper, it is important to mention
that experiments performed in quasi two-dimensional geometries, in wood 
\cite{wood-rou} or paper \cite{paper-rou}, 
yield a self-affine exponent close to the minimum energy surface result 
$\zeta_{loc}=2/3$. 

Scaling is also observed in acoustic emission experiments, 
where the distribution of pulses decays as a power law over several decades.
Experimental observations have been reported for several 
materials such as wood \cite{ciliberto}, cellular glass
\cite{strauven}, concrete \cite{ae} and paper \cite{paper},
but universality in the scaling exponents does not appear to be present.

The experimental observation of scaling behavior suggests 
an interpretation in terms of critical phenomena,
but a complete theoretical explanation has not been found.
The motion of a crack front has been modeled as a deformable line
pushed by the external stress through a random toughness
landscape.  Deformation of the crack surface is caused by disorder
and opposed by the elastic stresses. In certain conditions, the
problem can be directly related to models and theories of interface
depinning in random media and the roughness exponent computed
by numerical simulations and renormalization group calculations 
\cite{natt,nf}. Unfortunately, the numerical agreement between this 
theoretical approach and experiments is quite poor \cite{ram1,schmit2,ram2}.

One aspect missing from the crack line model is the nucleation
of voids in front of the main crack, an effect that has been shown 
to occur experimentally \cite{cel-03}. In this perspective, disordered lattice models
appear to be more appropriate to describe the phenomenon. In these models
the elastic medium is described by a network of  springs with random 
failure thresholds. In the simplest approximation of a scalar displacement,
one recovers the random fuse model (RFM) where a lattice of fuses with
random threshold are subject to an increasing external voltage
\cite{deArcangelis85,kahng88,delaplace,deArcangelis89,nukalajpamg}.
The model has been numerically simulated to obtain the  roughness of 
the fracture surface in two \cite{hansen91b,sep-00} and three dimensions 
\cite{bat-98,rai-98}. The measured roughness exponents turn out to be similar
to the one describing a minimum energy surface (or a directed polymer in $d=2$)
suggesting that crack formation occurs by an optimization process. 

In addition, the fracture of the RFM is preceded by avalanches of failure  
events \cite{hansen,zrvs,alava}. These are reminiscent of the acoustic emission
activity observed in experiments. The distribution of avalanche sizes (i.e. the number
of bonds participating in an avalanche)  follows a power law. In previous simulations
the exponent resulted to be close to $\tau=5/2$ , the value expected in the 
fiber bundle model (FBM) \cite{dfbm,hansen1}. In the FBM, load is 
redistributed equally in all the  fibers, representing thus a sort of 
mean-field limit of the RFM \cite{zrvs}. 
The load transfer in the RFM is long-ranged and is thus possible that RFM and FBM
display universal behavior \cite{BAR-02}. An intermediate case is provided by FBM with
long-range (power law) load transfer \cite{HID-02}: the difference with the RFM lies
in the anisotropic current transfer function \cite{BAR-02}. 

Numerical simulation of fracture in the RFM is often
hampered by the high computational cost associated with solving a
new large set of linear equations every time a new lattice bond is
broken. Previously, this fact has restricted the simulations to smaller lattice sizes and fewer 
statistical sampling of data, thereby affecting the quality of the results.
Here, thanks to the new algorithm discussed in Ref.~\cite{nukalajpamg}, 
we report results of numerical simulations for large 
two-dimensional lattices (triangular and diamond) with extended 
statistics. We concentrate on the roughness of the final crack
and the avalanche statistics preceding failure.

Using local and global measurements for the roughness we find
that cracks in the RFM follow anomalous scaling \cite{anomalous}. The local 
roughness exponent is found to be in the range $\zeta_{loc}=0.70-0.75$ ,
while the global exponent falls in the range $\zeta=0.80-0.85$. 
Although the difference between $\zeta$ and $\zeta_{loc}$ is small
it appears to be systematic. The results are obtained using the local width 
and the power spectrum methods and appear to be universal with respect 
to the lattice type. As a further test, we compute the width distribution
that can be collapsed into a unique curve for different lattice sizes
and types \cite{W-dist}. 

Next, we consider the distribution of avalanche sizes. The avalanche
signal is not stationary and as the current is raised avalanches becomes
larger and larger. The last avalanches, producing the failure of the
sample, is typically much larger than the previous one and it follows
a normal distribution with a typical value scaling as $s_m\sim L^{1.4}$.
Preceding avalanches are distributed as a power law with a cutoff
increasing with the current. Integrating the distribution over 
all the values of the current, we find a power law up to a cutoff, 
scaling with the lattice size as $L^D$, where $D\simeq 1.18$
does not depend on the lattice type and is thus universal. The 
exponent describing the decay of the distribution is found instead
to differ for triangular and diamond (square lattice
with 45 degrees inclined bonds to the bus bars) lattices with a value which
is always larger than the FBM value $\tau=5/2$.

The paper is organized as follows: in section II we define the
model, in section III we report the results on the crack roughness
and section IV is devoted to the avalanche statistics and in
section V we conclude.

\section{the random fuse model}

In the RFM \cite{deArcangelis85}, the lattice is
initially fully intact with bonds having the same conductance and
random breaking thresholds $t$, uniformly distributed between 0 and
1. The burning of a fuse occurs irreversibly, whenever the electrical
current in the fuse exceeds breaking threshold $t$ of the
fuse. Periodic boundary conditions are imposed in the horizontal
direction to simulate an infinite system and a constant voltage
difference, $V$, is applied between the top and the bottom of lattice
system bus bars. Numerically, a unit voltage difference, $V = 1$, is
set between the bus bars and the Kirchhoff equations are solved to
determine the current flowing in each of the fuses. Subsequently, for
each fuse $j$, the ratio between the current $i_j$ and the breaking
threshold $t_j$ is evaluated, and the bond $j_c$ having the largest
value, $\mbox{max}_j \frac{i_j}{t_j}$, is irreversibly removed
(burnt).  The current is redistributed instantaneously after a fuse is
burnt implying that the current relaxation in the lattice system is
much faster than the breaking of a fuse.  Each time a fuse is burnt,
it is necessary to re-calculate the current redistribution in the
lattice to determine the subsequent breaking of a bond.  The process
of breaking of a bond, one at a time, is repeated until the lattice
system fails completely. At this point we analyze the morphology of the 
spanning crack.

The same breaking sequence is obtained by raising the voltage difference 
or the total current at an infinitesimal rate. Doing this one can
identify an avalanche as the set of fuses breaking between two successive
increases of the voltage (or the current). In this paper, we follow Ref.~\cite{zrvs} ,
considering only current driven avalanches. The avalanche size is defined
as the number of fuses in an avalanche.  

Simulations are performed on two dimensional triangular and diamond
lattices of linear sizes going from $L=16$ up to $L=1024$ (for the triangular lattice)
or up to $L=256$ (for the diamond lattice). The total number of bonds in
the lattice is given by $N=(3L+1)(L+1)$ for the triangular lattice and $N=2L(L+1)$ for
the diamond lattice. Several results discussed in the following sections could
only be obtained under an extensive statistical sampling. Due to numerical
limitations this could not be achieved for the largest lattice sizes.
Each numerical simulation was performed on a single
processor of {\it Eagle} (184 nodes with four 375 MHz Power3-II
processors) supercomputer at the Oak Ridge National
Laboratory. The statistically independent $N_{config}$ number
of configurations were simulated simultaneously on number of
processors available for computation (the actual values of $N_{config}$
is reported in Table 1 of Ref.~\cite{nsz}).

\section{crack roughness}

After the sample has failed we identify the final crack, an example of
which is reported in Fig.~\ref{fig:1}. The cracks typically display some
limited amount of dangling ends and overhangs. We remove them and obtain
a single valued crack line $y_x$, where the values of $x \in [0,L]$ depend on the 
underlying lattice topology. 
Several methods have been devised to characterize the roughness of
an interface and their reliability has been tested against synthetic
data \cite{sch-95}. If the interface is self-affine all the methods should yield
the same result in the limit of large samples. For instance the local width, 
$w(l)\equiv \langle \sum_x (y_x- (1/l)\sum_X y_X)^2 \rangle^{1/2}$,
where the sums are restricted to regions of length $l$ and the average
is over different realizations, should scale as $w(l) \sim l^\zeta$
for $l \ll L$ and should saturate to a value $W=w(L) \sim L^\zeta$ corresponding
to the global width. The power spectrum 
$S(k)\equiv \langle \hat{y}_k \hat{y}_{-k} \rangle$, where 
$\hat{y}_k \equiv \sum_x y_x \exp i(2\pi xk/L)$, should decay as
$S(k) \sim k^{-(2\zeta+1)}$.

While numerical estimates with the two methods above could yield different
results, it is also possible that the scaling is anomalous\cite{anomalous}. This has been
observed not only in various growth models \cite{anomalous}
but also in fracture surfaces in granite
and wood samples\cite{exp-ano}. Anomalous scaling implies that the exponent describing the 
system size dependence of the
surface {\it differs} from the local exponent measured for a fixed system
size $L$. In particular, the local width scales as 
$w(l) \sim l^{\zeta_{loc}}L^{\zeta-\zeta_{loc}}$, so that the global 
roughness $W$ scales as $L^\zeta$ with $\zeta>\zeta_{loc}$. Consequently, the
power spectrum scales as $S(k) \sim k^{-(2\zeta_{loc}+1)}L^{2(\zeta-\zeta_{loc})}$.

Previous measurements of the crack roughness in the 
two-dimensional random fuse model have been obtained studying the global
roughness and anomalous roughness could not be detected. Here, thanks to the
improved statistics and system size range, we reveal clear indication of
anomalous scaling behavior. In Fig.~\ref{fig:2} we report the local width
for diamond and triangular lattices for different sizes $L$. The curves for
different system sizes are not overlapping even for $l \ll L$ as expected
for anomalous scaling. The global width scales with an exponent $\zeta =0.80 \pm 0.02$ and
$\zeta=0.83 \pm 0.02$ for diamond and triangular lattices respectively. On the
other hand the local width increases with a smaller exponent, that can be estimated
for the larger system sizes as $\zeta_{loc} \simeq 0.7$ for both lattices.
A more precise value of the exponents is obtained from the power
spectrum, which is expected to yield more precise estimates. 
Fig. \ref{fig:3} reports the data collapse of the power spectra for different
system sizes. The data are collapsed using $\zeta-\zeta_{loc}=0.1$ and
$\zeta-\zeta_{loc}=0.13$ for diamond and triangular lattices, respectively.
A fit of the power law decay of the spectrum yields instead $\zeta_{loc}=0.7$
and  $\zeta_{loc}=0.74$ for the two lattices, implying $\zeta=0.8$ and
$\zeta=0.87$. The results are close to the real space estimates and we
can attribute the differences to the bias associated to the methods employed \cite{sch-95}.

Although the value of $\zeta-\zeta_{loc}$ is small, it is significantly
larger than zero so that we would conclude that anomalous scaling
is present. While the local exponent is close to the directed polymer
value $\zeta=2/3$, the global value is much higher. In addition,
the presence of anomalous scaling would invalidate universality 
between directed polymers and fracture: directed
polymers should not display anomalous scaling. As for the question
of universality of the random fuse model crack roughness exponents,
the  values measured above are quite close to each other and the differences
could be due to size effects. In order to have a further confirmation
of this, we have analyzed the distribution $P(W)$ of the crack global width.
This distribution has been measured for various interfaces in models and
experiments and typically rescales as \cite{W-dist}
\begin{equation}
P(W)=P(W/\langle W \rangle)/\langle W \rangle,
\label{eq:widthdist}
\end{equation}
where $\langle W \rangle \sim L^\zeta$ is the average global width.
The crack width distribution has been measured for the random fuse model
with limited statistical sampling. We show in Fig.~\ref{fig:4} that the
distributions can be collapsed well using Eq.~\ref{eq:widthdist}
for diamond and triangular lattices.
The plot in Fig.~\ref{fig:5} shows that the collapsed distribution
for the two lattices superimpose, which we consider as a further indication
of universality. Finally, the width distributions are well fit by
a log-normal distribution as shown in Fig.~\ref{fig:5}.

\section{avalanches}

The qualitative behavior of the avalanche statistics is well understood in FBM, which 
can be solved exactly representing a mean-field version of the RFM \cite{zrvs}. 
The FBM can be formulated  as a parallel set of fuses, with random breaking threshold, under a
constant applied current $I$. Thus each
fuse carry the same current $f_i=I/n$, where $n$ is the number of intact
fuses. The FBM has been solved exactly ant it is known that there is
a critical value $I=I_c$ at which the bundle fails through a macroscopic 
avalanche. For $I<I_c$ fuses burn in smaller avalanches,
whose sizes are distributed as 
\begin{equation}
p(s,I^*)= s^{-\gamma}h(-s/s^*),\label{eq:binsize2}
\end{equation}
with $\gamma=3/2$, and $h(x)$ is a cutoff function. 
The cutoff size $s^*$ increases with the current  and close to $I_c$ 
diverges as $s^* \sim (I_c-I)^{1/\sigma}$ with $\sigma=1$.
One can then integrate the distribution over all the values of the
current, obtaining an $P(s) \sim s^{-\tau}$ with $\tau=\gamma+\sigma=5/2$.

Here we study the statistical properties of the avalanches in the RFM.
We can use the scaling laws established for the FBM as a reference, with
additional complications due to finite size effects.
In Fig.~\ref{fig:6} we report the integrated avalanche distribution
obtained for different lattice sizes. We observe a power law decay culminating
with a peak at large avalanche sizes. As in the FBM, the peak is due to the last catastrophic
event which can thus be considered as an outlier and analyzed separately.
When the last avalanche is removed from the distribution the peak disappears
(see Fig.~\ref{fig:6}). 

The avalanche size distribution, once the last event is excluded, is a power
law followed by an exponential cutoff at large avalanche sizes. The cutoff size $s_0$ 
is increasing with the lattice size, so that we can describe the distribution
by a scaling form 
\begin{equation}
P(s,L)=s^{-\tau} g(s/L^D),
\end{equation}
where $D$ represents the fractal dimension of the avalanches. To take into
account the different lattice geometries, it is convenient to express
scaling plots in terms of $N$ rather than $L$
\begin{equation}
P(s,N)=s^{-\tau} g(s/N^{D/2}).
\label{eq:psd}
\end{equation}

A powerful method to test these scaling laws, extracting $\tau$ and $D$, 
is provided by the moment analysis \cite{moments}. We compute the $q$th 
moment of the distribution
$M_q\equiv \langle s^q \rangle$ and plot it as a function of $N$. This defines
an exponent $\sigma_q$ as $M_q \sim N^{\sigma_q}$. If the data follow Eq.~\ref{eq:psd}
then $\sigma_q=0$ for $q<\tau-1$ and $\sigma_q=D(q+1-\tau)/2$ for $q>\tau-1$.
In order to measure $\sigma_q$, we consider lattice sizes from $L=16$ to $L=128$
since the statistical sampling for larger sizes is not adequate to estimate
correctly the cutoff $s_0$.
The data displayed in Fig.~\ref{fig:7} show that indeed $\sigma_q$ is linear in $q$ at large $q$
and vanishes for small $q$. The curves for triangular and diamond lattice do
not coincide: the two lines are parallel, indicating that $D$ is the same, but the
intersection with the $x$ axis differs. By a linear fit we obtain $\tau=2.75$ and
$D/2=0.59$ for diamond lattices and $\tau=3.05$ and $D/2=0.585$ for triangular
lattices. To confirm these results we perform a data collapse using the estimated values
of the exponents and result is reported in Fig.~\ref{fig:8}. While the data collapse for
diamond lattice is nearly perfect, some deviations are noticeable 
for the triangular lattice.

From the analysis discussed above, we would conclude that the
avalanche fractal $D$ dimension is universal, but a significant difference
is present for the exponent $\tau$. This difference could be due to lattice 
finite size effect as we will discuss later. In addition, the value of $\tau$ appears to 
be larger than the mean-field result $\tau=5/2$
obtained in the FBM. On the basis of less accurate results,
it was conjectured in Ref.~\cite{zrvs} that avalanches in the random fuse model are
ruled by mean-field theory. The present results seem to rule out this possibility.

So far we have considered avalanche statistics integrating the distribution
over all the values of the current. We have noticed, however, that the avalanche signal
is not stationary: as the current increases so does the avalanche size. In particular,
the last avalanche is much larger than the others. Its typical size grows as
$s_m \sim N^b$, with $b\simeq 0.7$ see Fig. 4 of Ref.~\cite{nsz} 
($s_m$ is referred as $n_f-n_p$ in that paper), while the distribution is approximately 
Gaussian as shown from the data collapse reported in 
Fig.~\ref{fig:9}

In Fig.~\ref{fig:10} we report the distribution of avalanche sizes sampled at different
values of the current $I$. For each sample, we normalize the current by its peak value $I_c$
and divide the $I^*=I/I_c$ axis into 20 bins. We then compute the avalanche size distribution
$p(s,I^*)$ for each bin and average over different realizations of the disorder.
In Fig.~\ref{fig:10} we report this distribution for a diamond lattice of size $L=128$. 
The distribution follows a law of the type
\begin{equation}
p(s,I^*)= s^{-\gamma}\exp(-s/s^*),\label{eq:binsize}
\end{equation}
with $\gamma \simeq 1.9$, while in the FBM $\gamma=3/2$.

In order to extract the dependence of the cutoff $s^*$ on $I^*$, 
we compute the second moment of the distribution $\langle s^2 \rangle$.
According to Eq.~\ref{eq:binsize}, this should scale as 
$\langle s^2 \rangle = (s^*)^{3-\gamma}$. Assuming that for large systems
$s^* \sim (1-I^*)^{-1/\sigma}$ (in the FBM this holds with
$\sigma=1$), we expect that the singularity is rounded at small $L$ as
\begin{equation}
s^* \sim \frac{L^D}{(1-I^*)^{1/\sigma}L^D+C},
\label{eq:fss}
\end{equation}
where $C$ is a constant. The second moment can be collapsed
very well under this finite size scaling assumption with $1/\sigma=1.4$
and $D=1.18$ as shown in Fig.~\ref{fig:11} for the diamond lattice. 
The data collapse is consistent with
the finite size scaling of the integrated distribution with a cutoff
increasing as $s_0 \sim L^D$. In fact integrating Eq.~\ref{eq:binsize}
we obtain
\begin{equation}
P(s,L) \sim s^{-(\gamma+\sigma)} \exp[-sC/L^D].  
\end{equation}
which implies $\tau=\gamma+\sigma$. Using the estimated data we would
obtain $\gamma+\sigma\simeq 2.6$ in reasonable agreement with the integrated
distribution result $\tau=2.75$.

We have performed the same analysis for the triangular lattice,
where we find similar scaling laws with $\gamma\simeq 2$ and $\sigma=1.3$.
This would give $\tau=2.7$ that is quite off from the integrated
distribution result $\tau=3.05$. These variations could indicate some
systematic error present in the triangular lattice results. We notice
that while in the diamond lattice, at the beginning of the simulation, 
all fuses carry the same current, in the triangular lattice only two thirds
of the fuses carry a current. As fuses break the current is redistributed becoming
inhomogeneous so that at breakdown this lattice effect should not be visible. 
In fact scaling exponents computed at failure, like the roughness exponent
or the avalanche cutoff, do not depend on the lattice type.
On the other hand, the integrated avalanche distribution is affected
by the entire rupture process and the estimated exponent could thus
be biased.

\section{conclusions}

In this paper we have revised some statistical properties of fracture in
the random fuse model using an improved statistical sampling and larger
lattices than what previously done in the past. We have analyzed the roughness
of the final crack for diamond and triangular lattices. The local roughness
exponent is found to be $\zeta_{loc}=0.72 \pm 0.03$ and appears to be different
from the global roughness exponent which turns out to be $\zeta =0.83 \pm 0.04$.
These results have been obtained from the local width and the power spectrum methods
and the error bars above merely represent the spread of the estimated exponents using
various methods and lattice types. The data suggest that anomalous scaling is present,
as already found in fracture experiments \cite{exp-ano}. 
The numerical value of the local exponent
is in reasonable agreement with the experiments on quasi two-dimensional materials
\cite{wood-rou,paper-rou}. As a further test for universality, we have also evaluated
the width distribution \cite{W-dist} that can be collapsed into a single curve
for different lattice sizes and types. From the theoretical point of view,
our results seem to exclude the minimum energy surface exponent of $\zeta=2/3$.
While the local exponent is close to that value, the global exponent is definitely
higher. In addition anomalous scaling is not expected for that model.
Thus the origin of measured roughness exponents and its theoretical explanation 
remains still open.

We have also analyzed the scaling of failure precursors, computing the distribution
of avalanche sizes. The extensive statistical sampling employed allowed us
to observe a power law decay up to a cutoff, which was not visible in
previous simulations \cite{hansen,zrvs}. The cutoff size is found to increase
with the lattice size as $s_0\sim L^D$, where the exponent $D\simeq 1.18$
depends very little on the lattice size. It is interesting to notice that for
self-affine lines of roughness, $\zeta$, one expects a fractal dimension $D=2-\zeta$ \cite{man2}.
If we plug into this expression the global roughness results obtained above for
the final crack, we  obtain $D\simeq 1.13-1.20$. This could imply that the
geometrical properties of the precursors are the same as that of the final crack. 
On the other hand, the exponent of the avalanche size distribution displays significant
variations with the lattice type (i.e. $\tau=2.75$ and $\tau=3.05$ for diamond and triangular lattices 
respectively) and is significantly different from the mean-field
result $\tau=5/2$ that was conjectured to be valid in \cite{zrvs}.

The integrated avalanche distribution is due to the convolution of the avalanche
distribution measured at different values of the current. We have shown that
the non-integrated distribution is given by a power law with an exponential
cutoff that increases with the current. The combined analysis of the distribution
with respect to current and lattice size can be performed using finite size
scaling. The behavior of the model is similar to the FBM, as 
noticed in Ref.~\cite{zrvs}, but the numerical
values of the exponents change. For the diamond lattice we estimate
$\gamma=1.9$ and $\sigma = 1.4$, while the FBM yields
$\gamma=3/2$ and $\sigma=1$. Similar results hold for the triangular lattice
although the scaling there appears to be less clear.

It would be interesting to understand these results theoretically 
by the renormalization group, using the 
mean-field theory as a reference. Steps in this direction
have been made in Ref.~\cite{BAR-02} but the complicated (dipolar)
structure of the current redistribution function makes the problem
very hard to deal with. Long-range interactions appear to be crucial
in the appearence of scaling behavior, since local fracture models
yield abrupt failure without large precursors \cite{KUN-00}. 
A similar scenario is characteristic of first-order phase transitions occuring 
close to a spinodal. In that case spinodal scaling is only seen in mean-field or with long range
interaction \cite{spinodal}. The fact that the exponents deviate from
mean-field ones, however, calls for a more detailed understanding of
the origin of scaling in the random fuse model.

\begin{widetext}

\begin{figure}[hbtp]
\centerline{\psfig{file=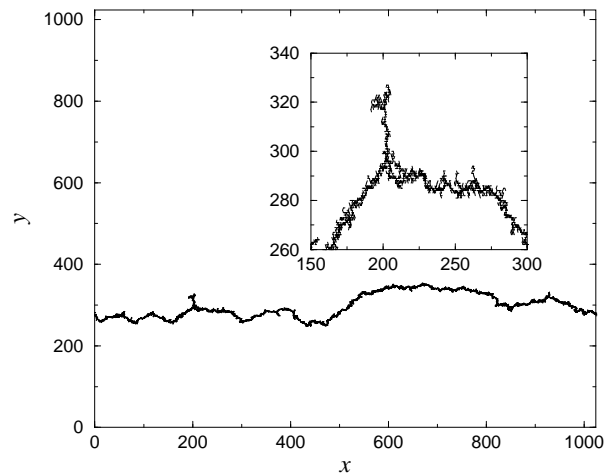,width=8cm,clip=!}}
\caption{The final crack in a triangular lattice of size $L=1024$ (a detail is
shown in the inset). The crack displays some dangling ends and overhangs that are
removed before performing the analysis.}
\label{fig:1}
\end{figure}

\begin{figure}[hbtp]
\centerline{\psfig{file=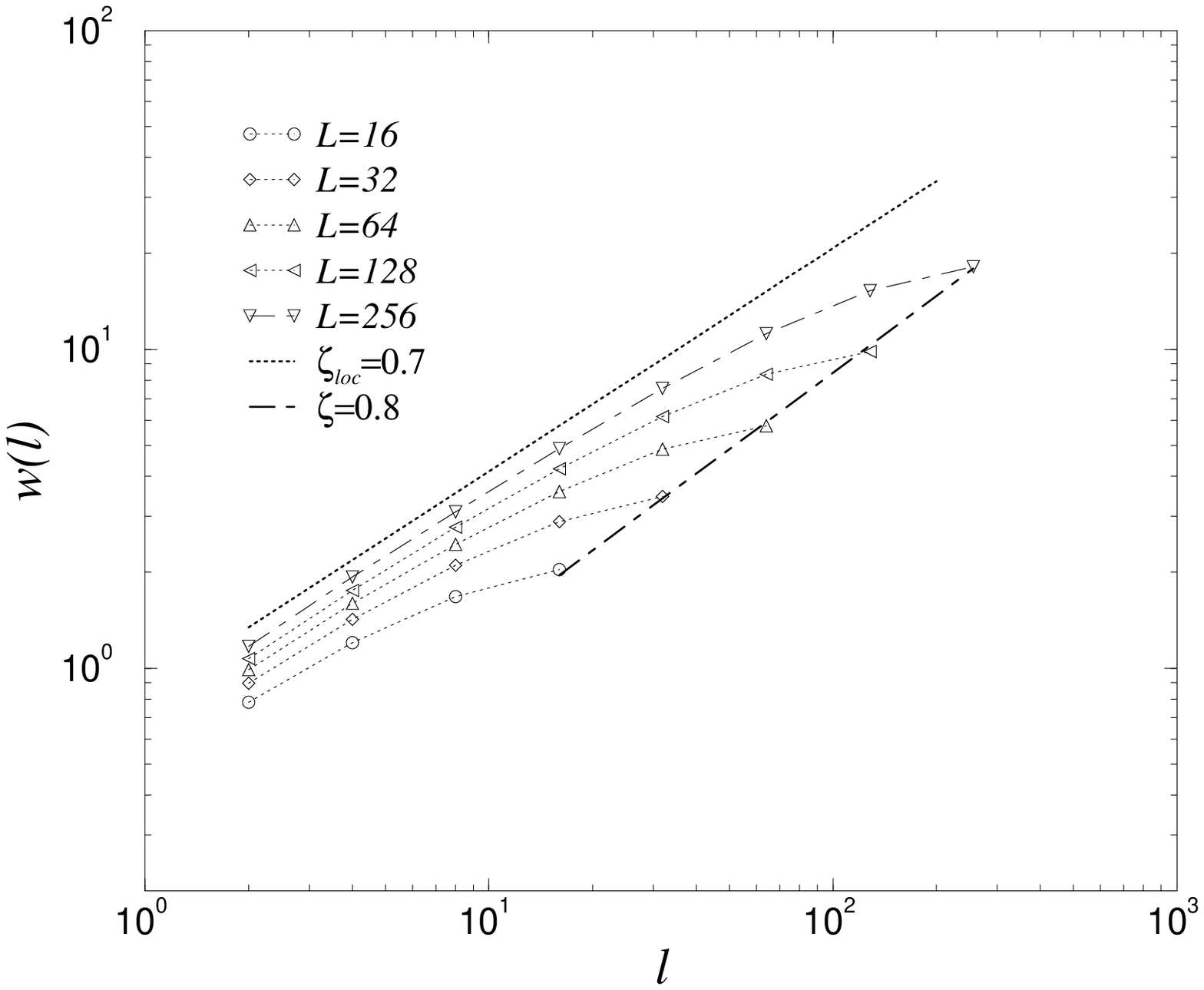,width=8cm,clip=!},
\psfig{file=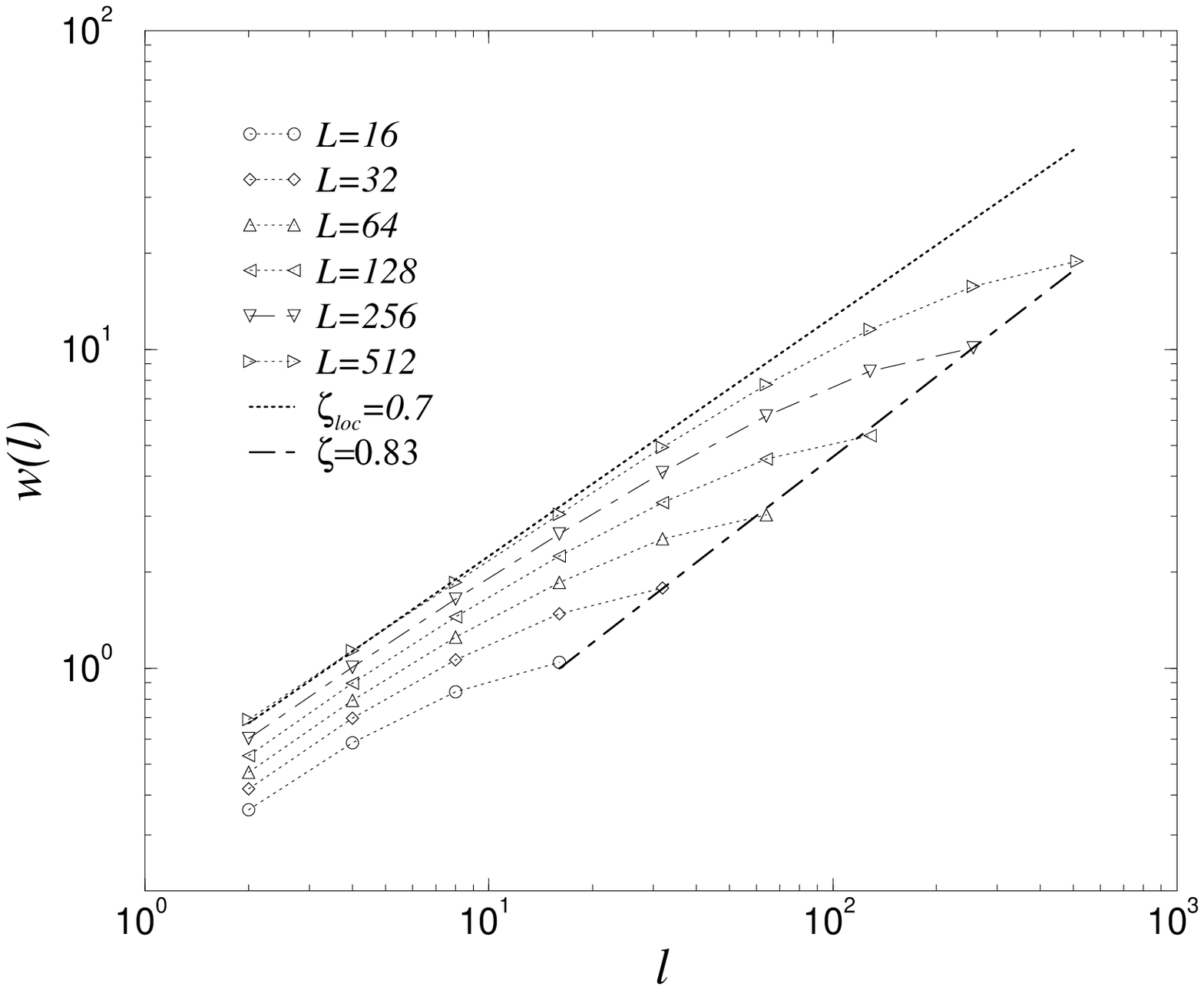,width=8cm,clip=!}}
\caption{The local width $w(l)$ of the crack for different lattice sizes in log-log scale. 
A line with the local exponent $\zeta_{loc}=0.7$ is plotted for reference.
The global width displays an exponent $\zeta>\zeta_{loc}$. Data are shown
for diamond (left) and triangular (right) lattices.}
\label{fig:2}
\end{figure}

\begin{figure}[hbtp]
\centerline{\psfig{file=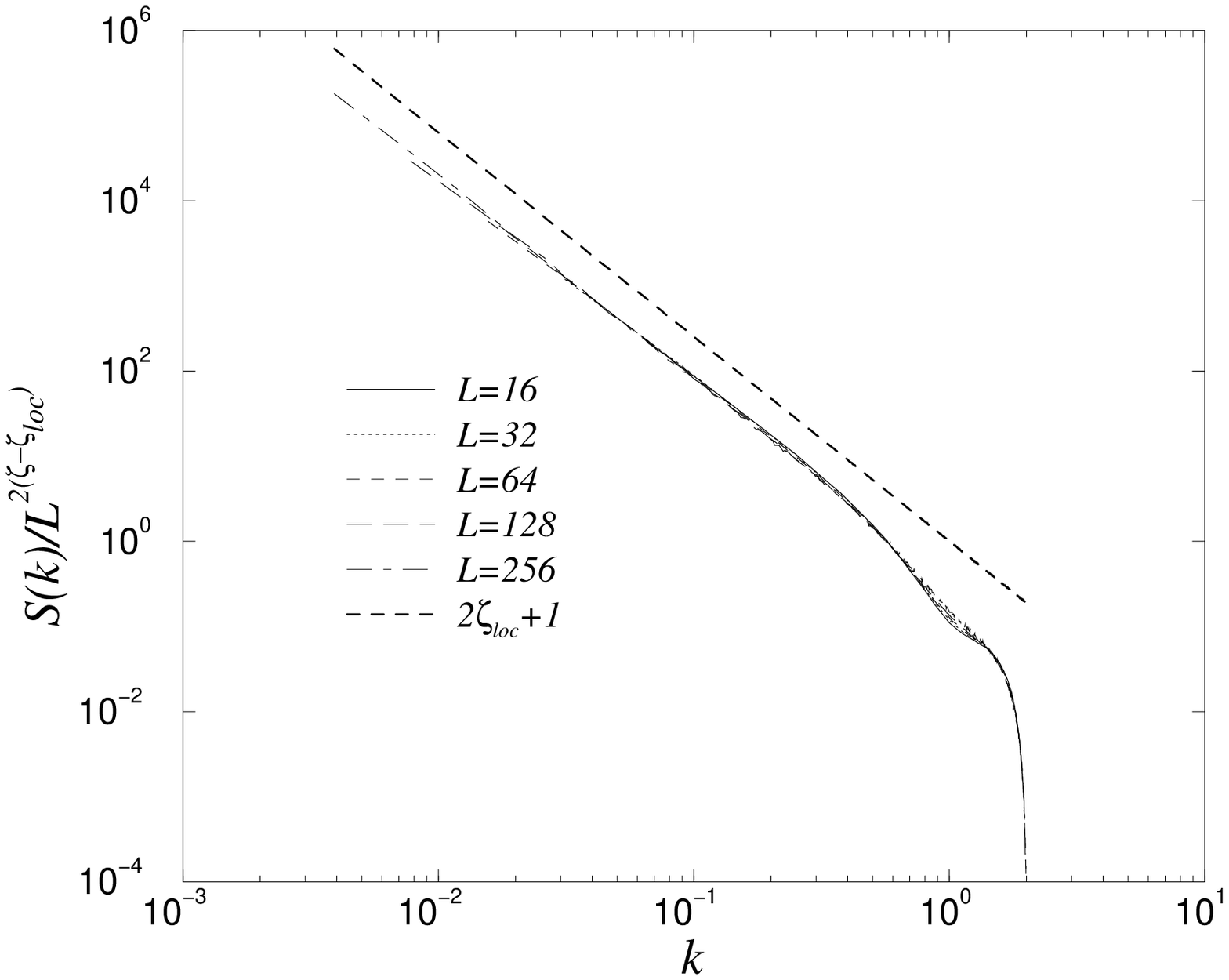,width=8cm,clip=!}\psfig{file=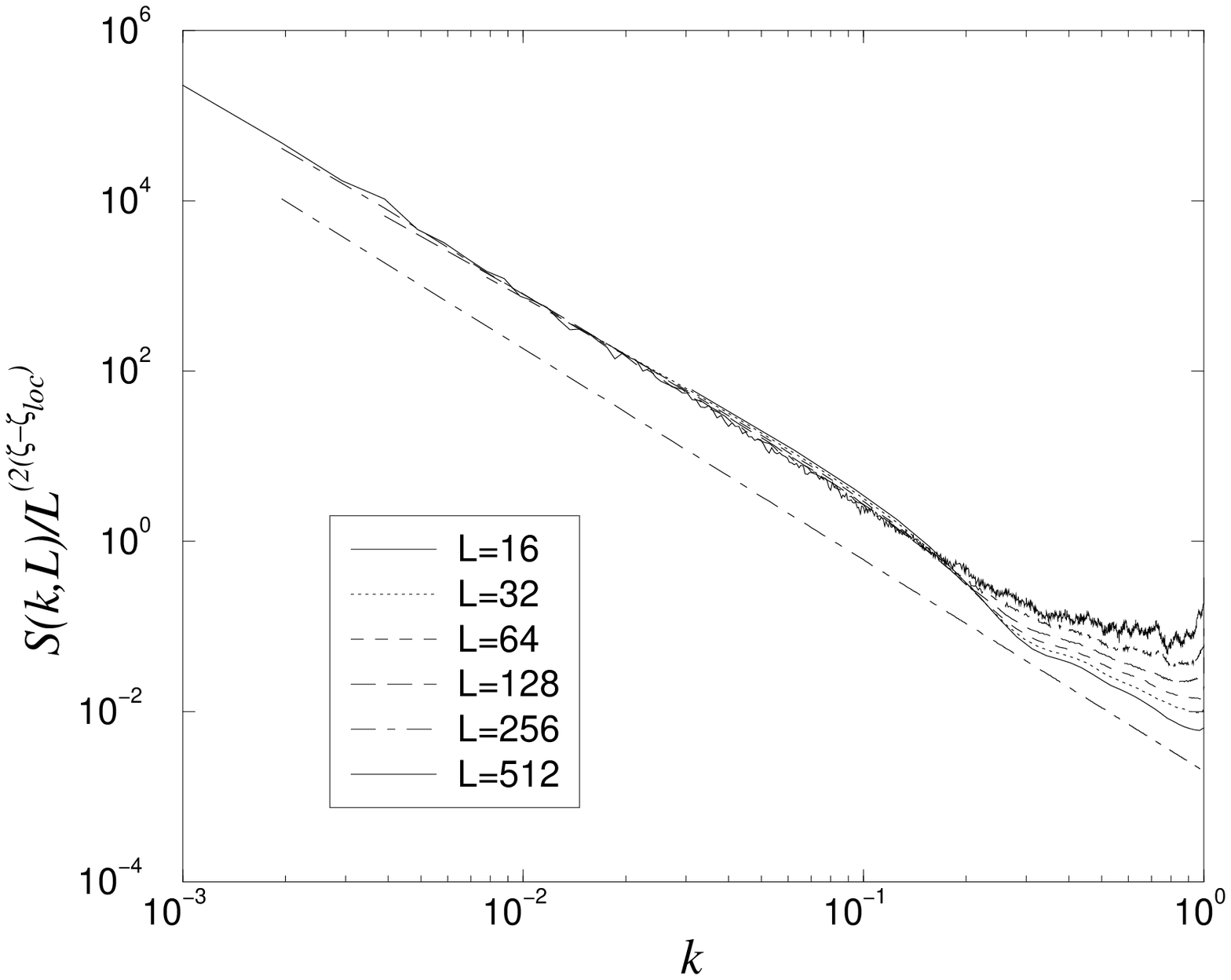,width=8cm,clip=!}}
\caption{The power spectrum of the crack $S(k,L)$ for different lattice sizes in log-log scale.  
The slope defines the local exponent as $-(2\zeta_{loc}+1)$. The spectra for all of the 
different lattice sizes can be collapsed indicating anomalous scaling. Data are shown
for diamond (left) and triangular (right) lattices. }
\label{fig:3}
\end{figure}

\begin{figure}[hbtp]
\centerline{\psfig{file=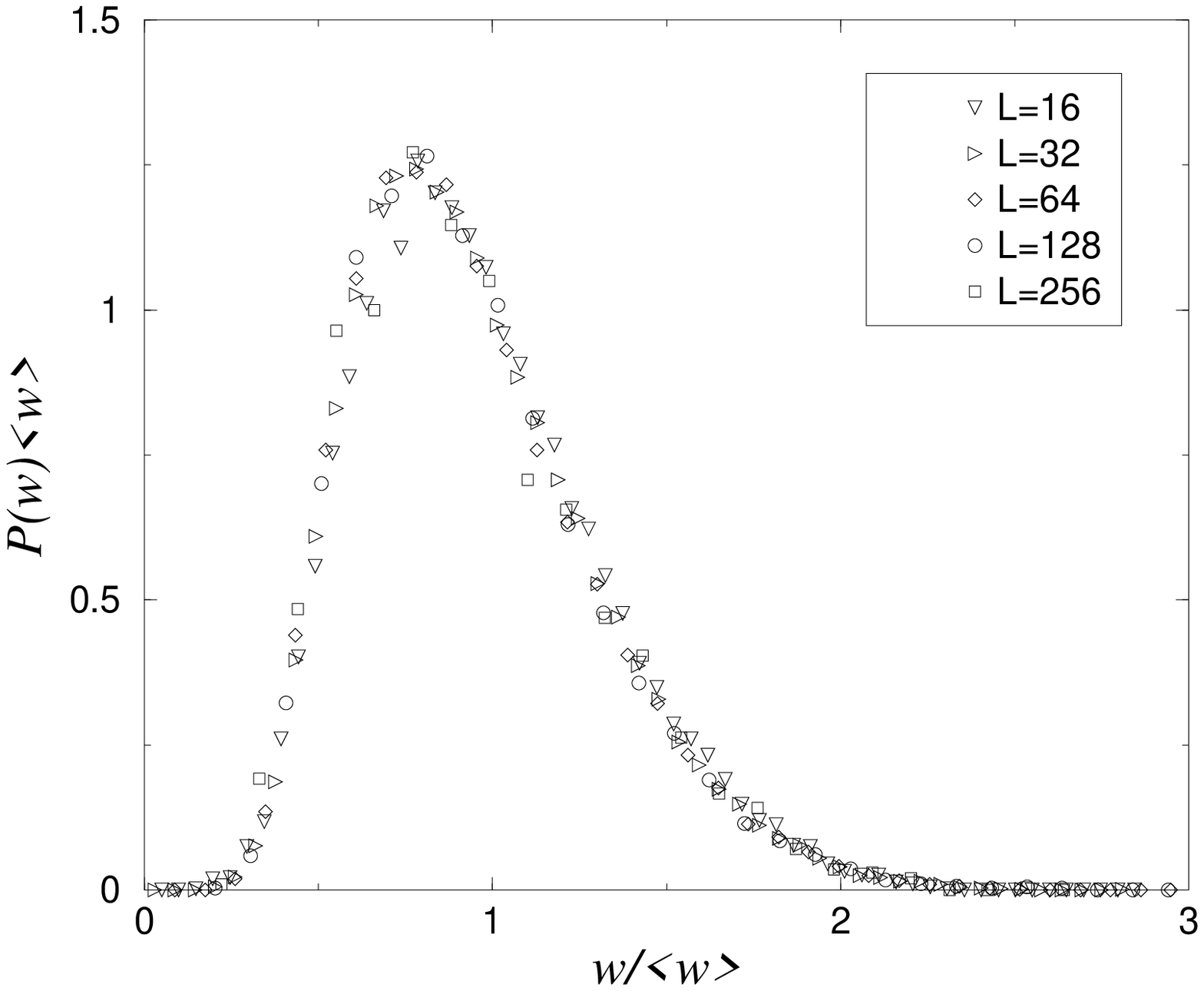,width=8cm,clip=!}\psfig{file=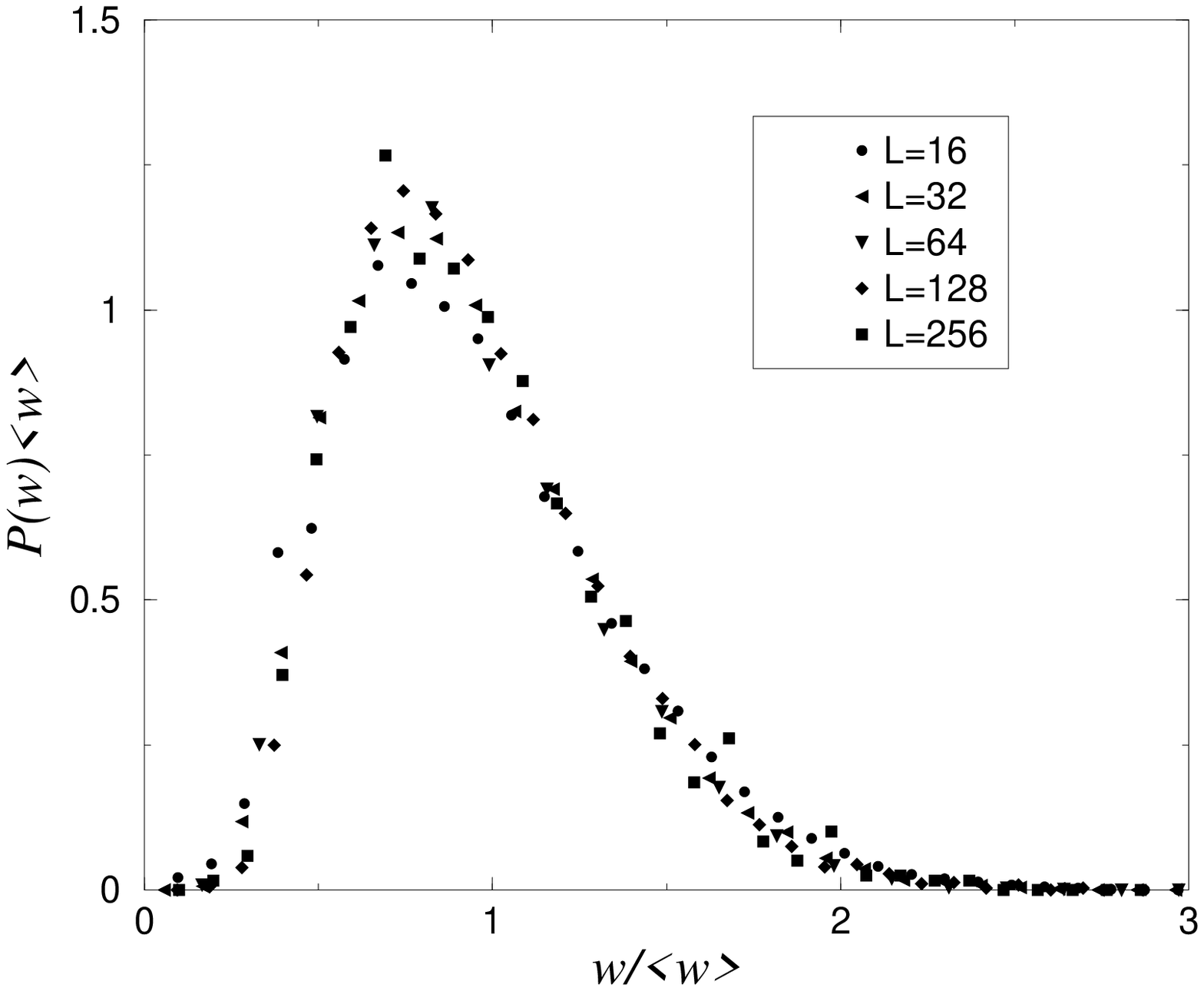,width=8cm,clip=!}}
\caption{The distribution of crack width for different lattice sizes can be
collapsed using their average value. Data are shown
for diamond (left) and triangular (right) lattices. }
\label{fig:4}
\end{figure}

\begin{figure}[hbtp]
\centerline{\psfig{file=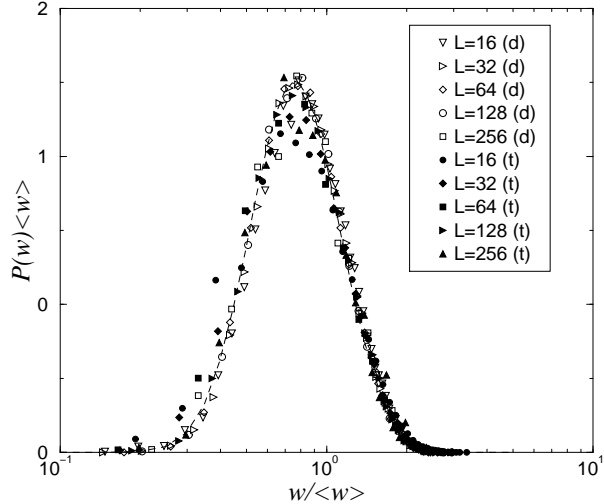,width=8cm,clip=!}}
\caption{The distribution of crack width is universal for diamond and triangular
lattices since all the curves can be collapsed together. 
A fit with a lognormal distribution is shown by a dashed line.}
\label{fig:5}
\end{figure}

\begin{figure}[hbtp]
\centerline{\psfig{file=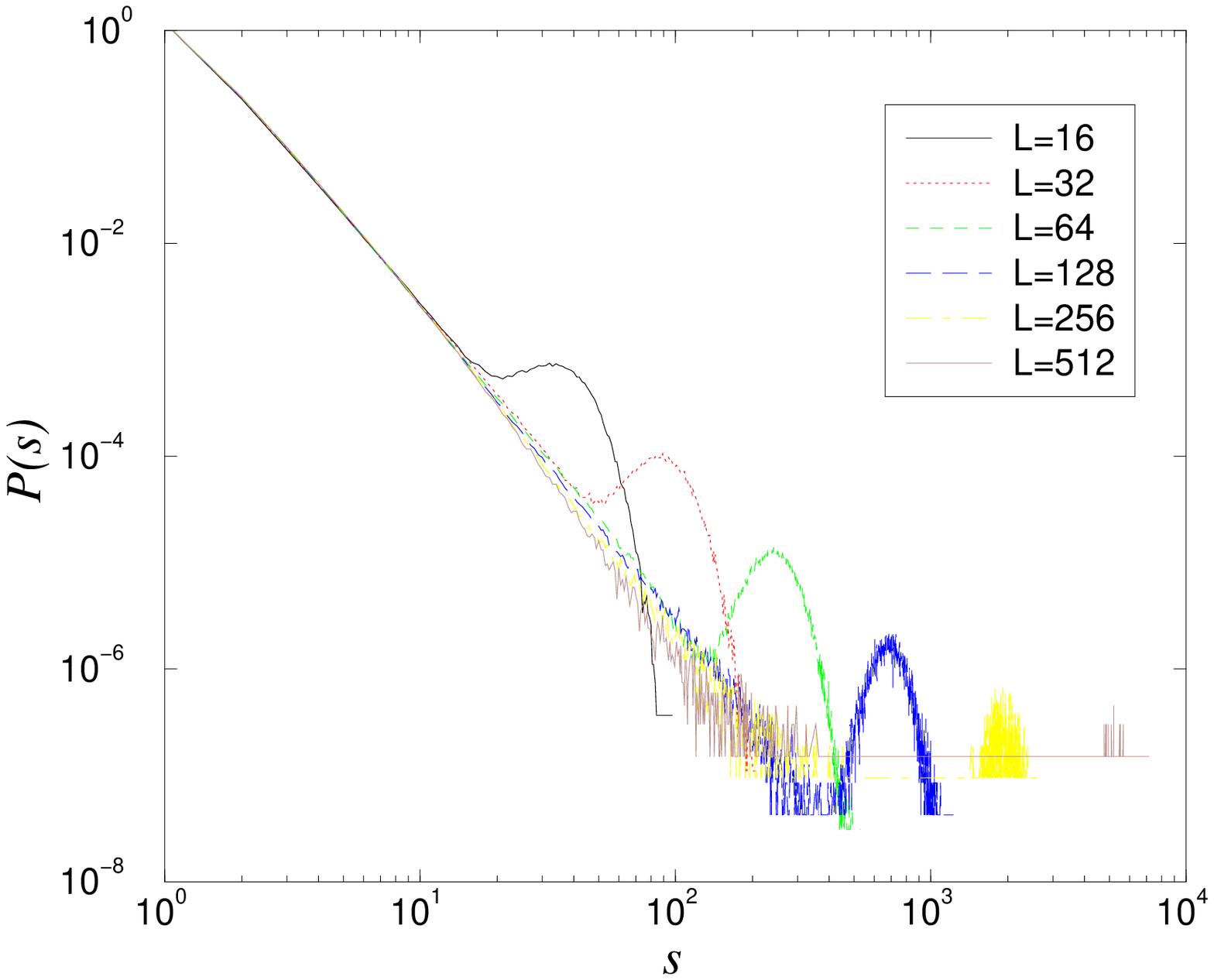,width=8cm,clip=!},
\psfig{file=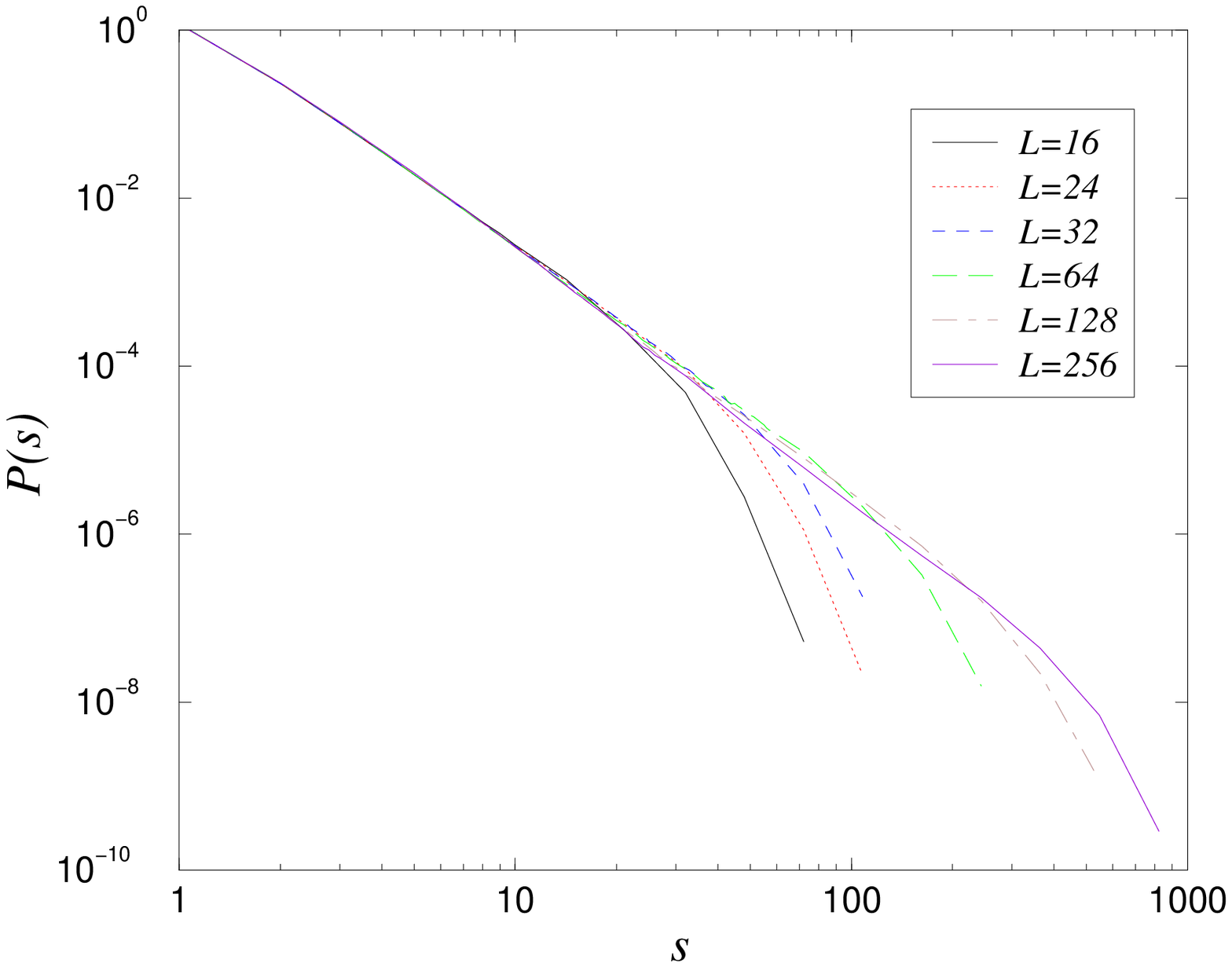,width=8cm,clip=!},}
\caption{The distribution of avalanche sizes for triangular lattices
of different sizes. The peak at large size is due to the last avalanche,
corresponding to catastrophic failure (right). On the left figure we show
the same distribution without the last event and with 
logarithmic bins.}
\label{fig:6}
\end{figure}

\begin{figure}[hbtp]
\centerline{\psfig{file=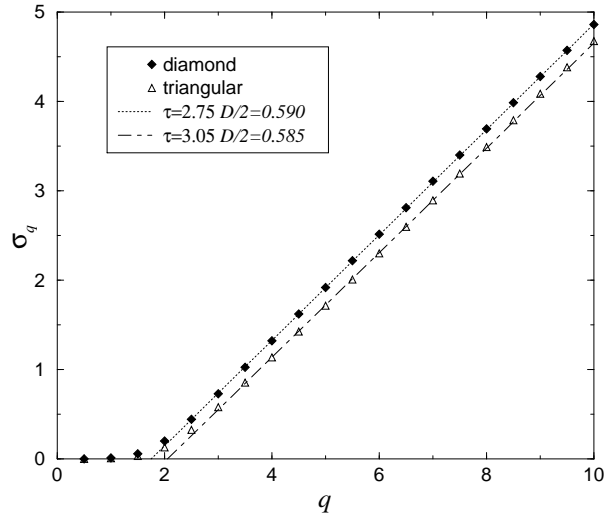,width=8cm,clip=!}}
\caption{The exponent $\sigma_q$ ruling the scaling of the $q$th moment for
triangular and diamond lattice. The shift in the lines indicates a difference
in the value of $\tau$.}
\label{fig:7}
\end{figure}

\begin{figure}[hbtp]
\centerline{\psfig{file=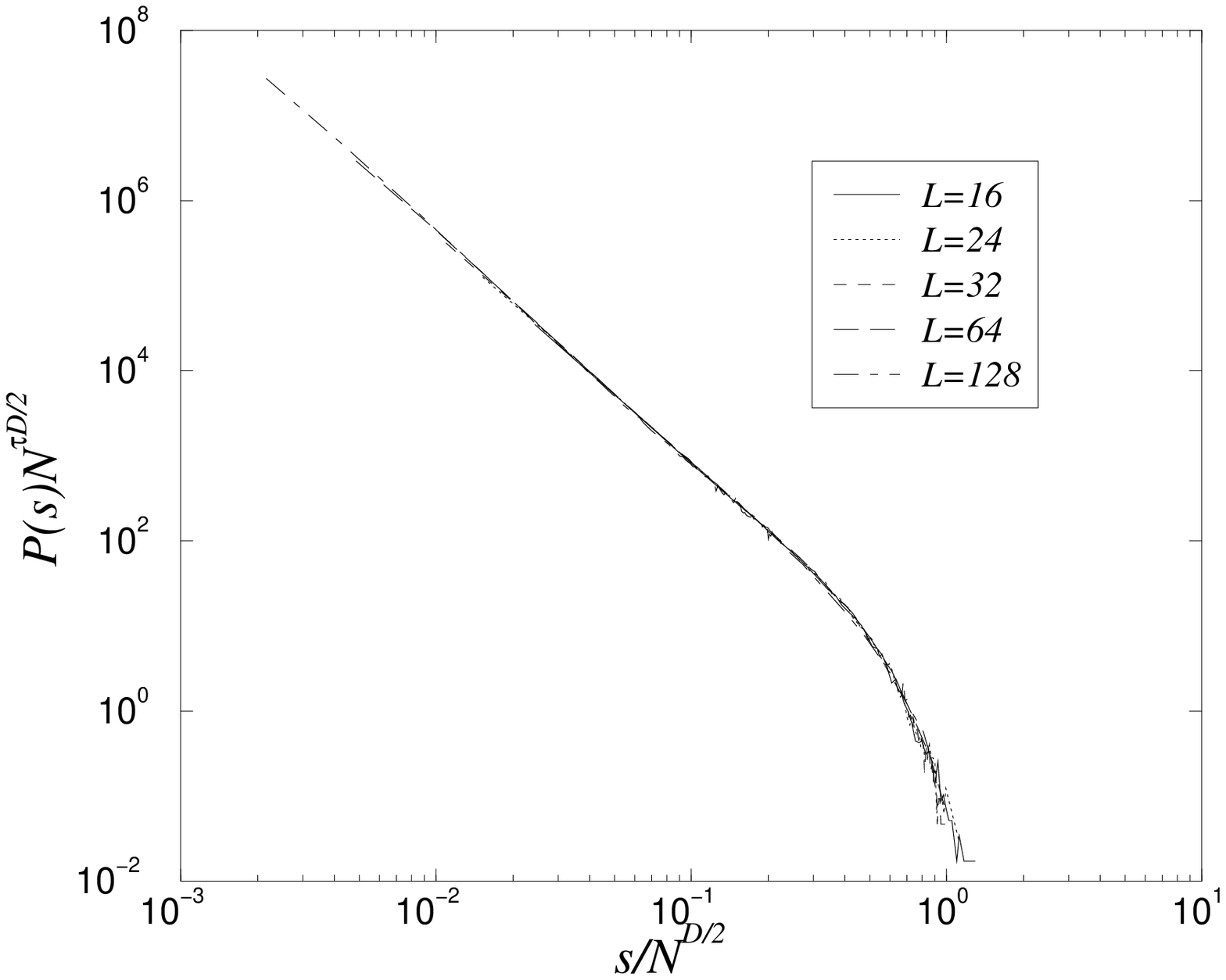,width=8cm,clip=!}
\psfig{file=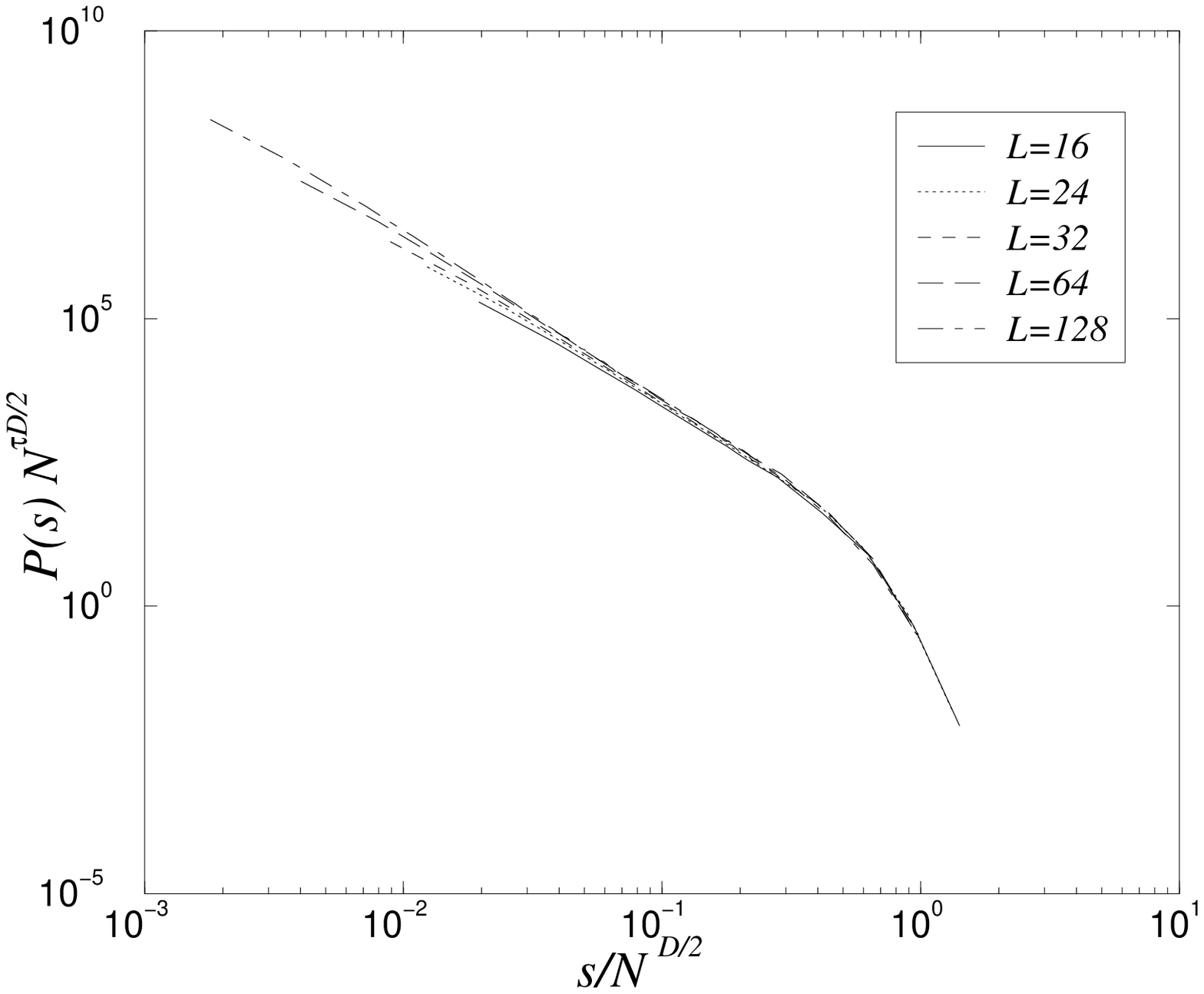,width=8cm,clip=!}}
\caption{Data collapse of the avalanche size distributions. The exponent used for
the collapse are $\tau=2.75$ and $D=1.18$ 
for the diamond lattice (left) and $\tau=3.05$ and $D=1.17$ for the
triangular (right) lattice.}
\label{fig:8}
\end{figure}

\begin{figure}[hbtp]
\centerline{\psfig{file=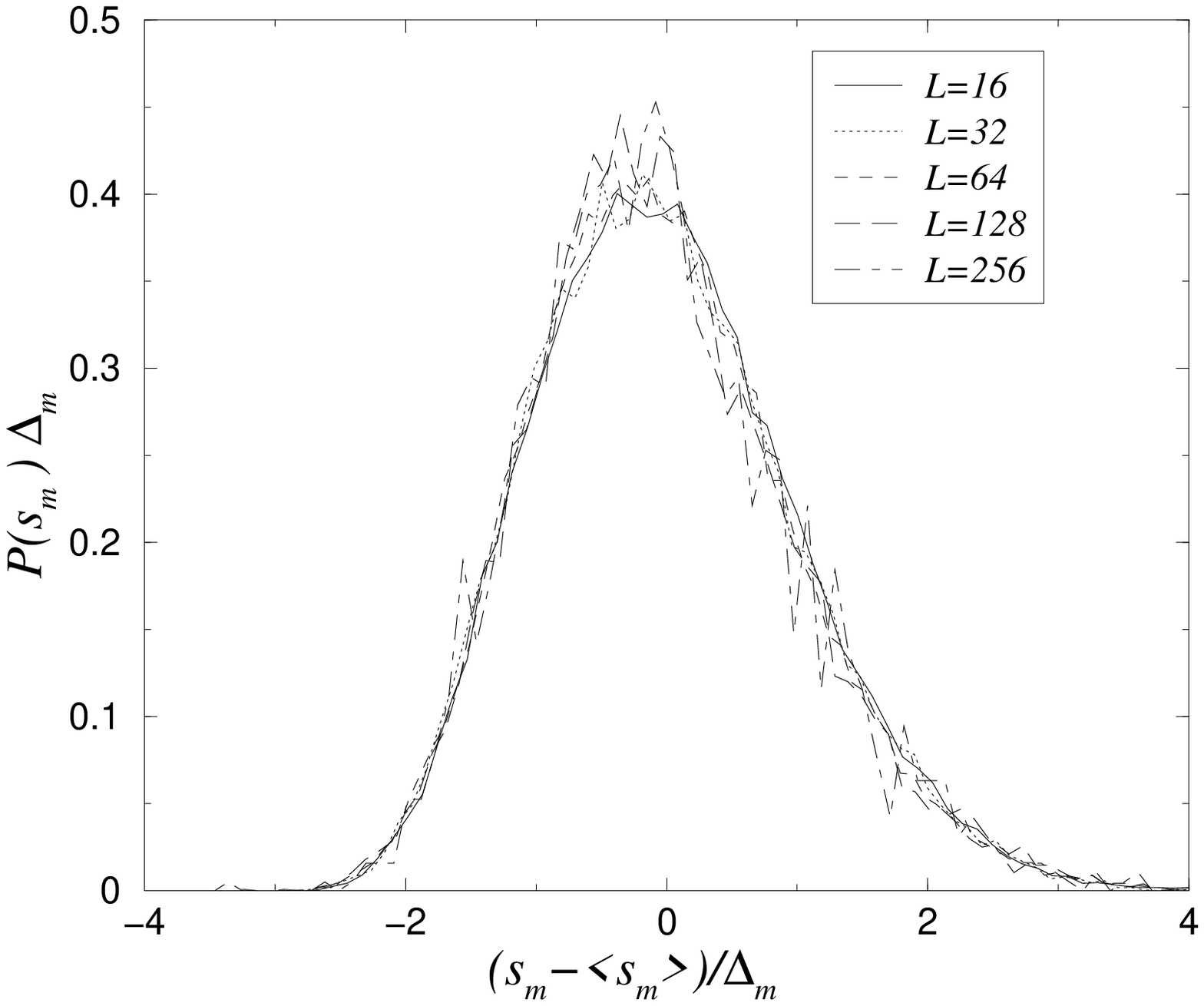,width=8cm,clip=!}
\psfig{file=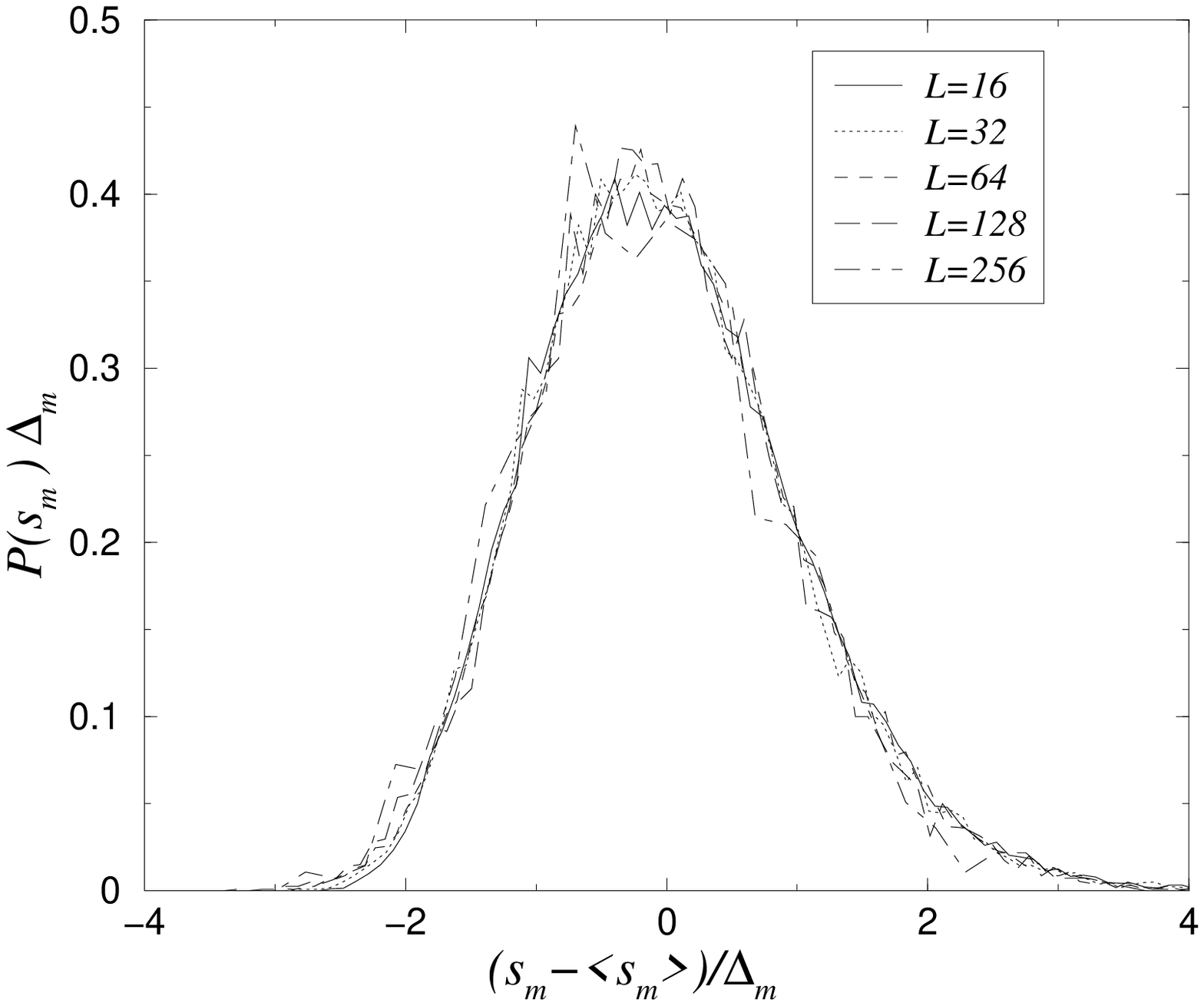,width=8cm,clip=!}}
\caption{Data collapse of the distribution of the last avalanche for
diamond (left) and triangular (right) lattice.}
\label{fig:9}
\end{figure}

\begin{figure}[hbtp]
\centerline{\psfig{file=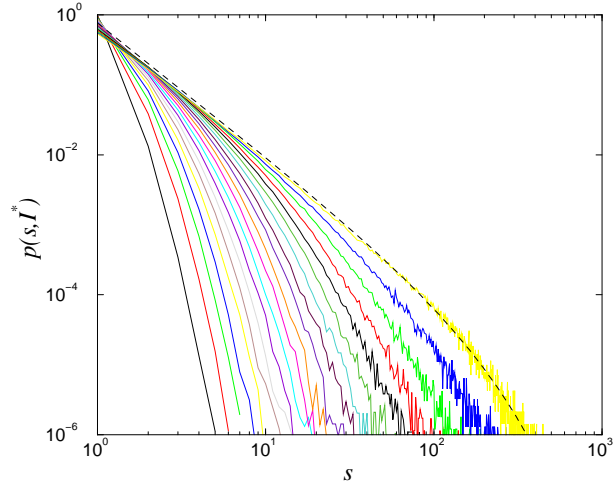,width=8cm,clip=!}}
\caption{The avalanche size distributions sampled over a small bin of the reduced current
$I^*$ for a diamond lattice of size $L=128$. 
The dashed line represents a fit according to Eq.~\protect\ref{eq:binsize} with $\gamma=1.9$.}
\label{fig:10}
\end{figure}

\begin{figure}[hbtp]
\centerline{\psfig{file=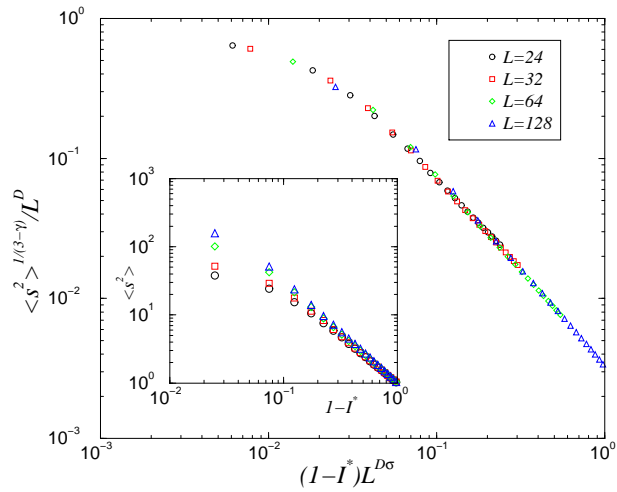,width=8cm,clip=!}}
\caption{The second moment of the avalanche size distribution 
as a function of the reduced current $1-I^*$ for diamond lattices of
different sizes (inset). 
The curves can be collapsed using the finite size scaling
assumption reported in Eq.~\protect\ref{eq:fss} with $\gamma=1.9$, $D=1.18$ and 
$1/\sigma=1.4$.}
\label{fig:11}
\end{figure}
\end{widetext}

\end{document}